\documentclass[aps, prb, twocolumn,superscriptaddress,amsmath,amssymb,reprint]{revtex4-2}
\usepackage{graphicx}
\usepackage{dcolumn}
\usepackage{bm}
\usepackage[dvipsnames]{xcolor}
\usepackage{physics}
\usepackage{pifont}
\usepackage{comment}
\usepackage{makecell}
\usepackage{svg}
\usepackage{amsmath}
\usepackage{lineno}

\begin{document}

\title{Spin-Chirality-Driven Bulk Photovoltaic Effect in van der Waals Magnet CrSBr}

\author{Dezhao \surname{Wu}}
\affiliation{State Key Laboratory of Low Dimensional Quantum Physics and Department of Physics, Tsinghua University, Beijing, 100084, China}

\author{Yong \surname{Xu}}
\email{yongxu@mail.tsinghua.edu.cn}
\affiliation{State Key Laboratory of Low Dimensional Quantum Physics and Department of Physics, Tsinghua University, Beijing, 100084, China}
\affiliation{Frontier Science Center for Quantum Information, Beijing, 100084, China}
\affiliation{RIKEN Center for Emergent Matter Science (CEMS), Wako, Saitama 351-0198, Japan}

\author{Meng \surname{Ye}}
\email{mye@gscaep.ac.cn}
\affiliation{Graduate School of China Academy of Engineering Physics, Beijing, 100193, China}

\author{Wenhui \surname{Duan}}
\affiliation{State Key Laboratory of Low Dimensional Quantum Physics and Department of Physics, Tsinghua University, Beijing, 100084, China}
\affiliation{Frontier Science Center for Quantum Information, Beijing, 100084, China}
\affiliation{Institute for Advanced Study, Tsinghua University, Beijing 100084, China}

\begin{abstract}
The bulk photovoltaic effect (BPVE) manifests with remarkable diversity in magnetic materials.
Here, we establish vector spin chirality as a tunable knob for generating an unconventional time-reversal-even magnetic BPVE, comprising the chiral shift current (CSC) and chiral injection current (CIC). Using bilayer antiferromagnetic (AFM) CrSBr as a prototype, we theoretically demonstrate the emergence of CSC and CIC. Compared with conventional photovoltaic currents arising from noncentrosymmetric crystal structures or collinear magnetic orderings, CSC and CIC not only possess comparable magnitudes but also exhibit exceptional tunability. Specifically, they can be switched on and off by magnetic-field-induced spin canting, reversed in direction upon canting-direction reversal, and continuously modulated in intensity via canting-angle variation. Furthermore, we reveal an unusual optical transition channel governing both currents in CrSBr. Our work establishes an unconventional magnetic BPVE with remarkable controllability, offering opportunities for optoelectronic and magnetic-sensing applications in noncollinear magnets.
\end{abstract}


\maketitle

\section*{Introduction}\label{Introduction}
The bulk photovoltaic effect (BPVE) is a second-order optical response that converts light into a steady-state electric current in single-phase materials lacking inversion ($\mathcal{P}$) symmetry, stimulating fundamental research and opening opportunities for optoelectronic applications based on quantum materials.
While the conventional time-reversal-even ($i$-type) BPVE arising from noncentrosymmetric crystal structures has been extensively investigated and measured in nonmagnetic materials \cite{sipe-2000-prb-over_band_gap_voltage,2018-Science-overcome_Shockley-Queisser_limit,2016-Shockley-Queisser-and-above-band-gap-photovoltage,Rappe-2012-prl-NSC_in_BaTiO3,Rappe-2012-prl-NSC_in_BaFeO3,Rappe-2016-prl-NSC_in_BiTeI&CsPbI3,Neaton-2017-prl-NSC_in_GeS,Moore-2017-NC-design_of_NSC,Qian-2019-SA-NSC&NIC_in_GeS,Huang-2023-prl-NSC_Heteronodal_Line_Systems,Kawasaki-2017-NC_NSC_experiment,Rappe-2017-npjCM_NSC_in_PbTiO3,Liu-2023-npjCM_2D_ferroelectrics,Yao-2024-NSC_in_PtBi2,Zhang-2024-nl-NSC_of_band_mixture,Zou-2024-NSC_in_GeTe}, the theoretical proposal and experimental discovery of the time-reversal-odd ($c$-type) magnetic BPVE arising from magnetic orderings significantly enrich the origins and mechanisms of the BPVE and are attracting increasing interest \cite{Yan-2019-NC-MIC_in_CrI3, Yan-2020-prr-carrier_lifetime, Qian-2020-npjCM-MIC&NSC_in_MnBi2Te4, Nagaosa-2020-prx, Watanabe-2021-prx-chiral_photocurrents, Ideue-2026-Nat-Mater-MIC_in_CrSBr, Wu-2026-arxiv-MSC_exp, Matsuda-2025-NC-BPVE_in_MoS2-CrPS4, Li-2020-prb-MIC_in_MnBi2Te4, Zhou-2022-nl-MIC&NSC_in_VSe2,Zhou-2023-prb-MIC_MnPSe3, Weng-2023-Quantum-Frontiers-MIC&MSC_in_EuSn2As2, Li-2024-JPCL-MIC&NSC_Nb3I8,Zhou-2026-AM-BPVE_review}. The $c$-type magnetic BPVE was first proposed and predicted in bilayer (BL) CrI$_3$, where the generated photocurrent can be uniquely switched by reversing the N\'eel vector \cite{Yan-2019-NC-MIC_in_CrI3}. Subsequently, the theory of magnetic BPVE was further developed through general analyses of responses enabled by time-reversal-symmetry ($\mathcal{T}$) breaking \cite{Yan-2020-prr-carrier_lifetime,Qian-2020-npjCM-MIC&NSC_in_MnBi2Te4}, a quantum-geometric framework \cite{Nagaosa-2020-prx}, and a systematic symmetry-based classification \cite{Watanabe-2021-prx-chiral_photocurrents}.
Very recently, switchable photocurrents upon reversal of the N\'eel vector have been experimentally observed in BL CrSBr \cite{Ideue-2026-Nat-Mater-MIC_in_CrSBr} and MnPSe$_3$ flakes \cite{Wu-2026-arxiv-MSC_exp}, representing a milestone in the exploration of the $c$-type magnetic BPVE.
However, recent experimental progress \cite{Ideue-2026-Nat-Mater-MIC_in_CrSBr,Wu-2026-arxiv-MSC_exp} alongside previous theoretical studies \cite{Yan-2019-NC-MIC_in_CrI3,Qian-2020-npjCM-MIC&NSC_in_MnBi2Te4,Li-2020-prb-MIC_in_MnBi2Te4,Zhou-2022-nl-MIC&NSC_in_VSe2,Zhou-2023-prb-MIC_MnPSe3,Weng-2023-Quantum-Frontiers-MIC&MSC_in_EuSn2As2,Li-2024-JPCL-MIC&NSC_Nb3I8} has primarily focused on photocurrents in collinear magnets, confining the focus to a limited material pool of two-dimensional (2D) antiferromagnetic (AFM) materials. Therefore, discovering fundamentally distinct mechanisms of BPVE in noncollinear magnets remains an urgent need, which will not only advance the study of the magnetic BPVE but also further broaden the functionalities of optoelectronic applications.

In contrast to collinear magnets, noncollinear magnets provide a promising avenue to realize a novel $i$-type magnetic BPVE. In such systems, the noncollinear arrangement of adjacent spins $\mathbf{S_A}$ and $\mathbf{S_B}$ allows the emergence of an $i$-type quantity, the vector spin chirality $\bm{\kappa} = \mathbf{S_A} \times \mathbf{S_B}$ \cite{chiral_Hall_2020_prl,chiral_Hall_2021_communications_physics,chiral_photocurrents_2023_APL,SA}, suggesting the existence of an unconventional spin-chirality-driven BPVE, which constitutes a distinct subclass of the magnetic BPVE yet is expected to be inherently $i$-type. Furthermore, for both the BPVE responses experimentally measured in spin-canted CrPS$_4$/MoS$_2$ heterostructures \cite{Matsuda-2025-NC-BPVE_in_MoS2-CrPS4} and theoretically predicted in spin-spiral NiI$_2$ \cite{Zhou-2024-prb-CSC_in_NiI2}, certain components show a sensitivity to noncollinear spin configurations, providing important motivation to clarify the role of $\bm{\kappa}$ in magnetic BPVE. Therefore, the exploration of the existence of the spin-chirality-driven BPVE through a comprehensive classification, alongside the elucidation of its unique properties and fundamental distinctions from the traditional BPVE, as well as the identification of suitable materials for experimental verification, remains urgently needed.

The 2D van der Waals layered semiconductor CrSBr, featuring an A-type AFM ordering \cite{2020_AM_CrSBr,zhu_2021_nanoletters_CrSBr,zhu_2021_nature-mater_CrSBr,xu_2022_nature-nanotech_CrSBr, zhu_2022_nature_CrSBr,wang_2022_acsnano_CrSBr,xu_2023_nature-nanotech_CrSBr_tunable_exciton-magnon_coupling, xu_2025_nature-nanotech_CrSBr_tunable_exciton-magnon_coupling, 2023_Nature_CrSBr_tunable_MO,2024-NC-CrSBr-doping,CrSBr_2023_1D_ACS-nano, zhu_2024_CrSBr_nanoletter_review}, provides an ideal platform for exploring the concept of the spin-chirality-driven BPVE.
First, CrSBr can host a nonvanishing $\bm{\kappa}$ via spin canting induced by a weak external magnetic field ($<$ 2\,T) \cite{2020_AM_CrSBr,zhu_2021_nanoletters_CrSBr,zhu_2021_nature-mater_CrSBr,xu_2022_nature-nanotech_CrSBr,zhu_2022_nature_CrSBr,wang_2022_acsnano_CrSBr, xu_2023_nature-nanotech_CrSBr_tunable_exciton-magnon_coupling, xu_2025_nature-nanotech_CrSBr_tunable_exciton-magnon_coupling, 2023_Nature_CrSBr_tunable_MO,2024-NC-CrSBr-doping,zhu_2024_CrSBr_nanoletter_review}. Second, it exhibits robust air stability, facile device fabrication, and a high N\'eel temperature \cite{zhu_2024_CrSBr_nanoletter_review}.
Third, its centrosymmetric crystal structure \cite{zhu_2021_nanoletters_CrSBr} rigorously precludes any conventional $i$-type BPVE of crystallographic origin, thereby rendering CrSBr a prototypical material for the isolated study of magnetic BPVE, especially the potential spin-chirality-driven BPVE.

In this work, employing symmetry analysis and first-principles calculations \cite{Wang_2017,Chen_2022,npj}, we uncover an unconventional magnetic BPVE driven by $\bm{\kappa}$, comprising the chiral shift current (CSC) and the chiral injection current (CIC). These spin-chirality-driven photocurrents are $i$-type, rendering them fundamentally distinct from the traditional $c$-type magnetic BPVE. Using BL AFM CrSBr as a prototype, we find that the CSC and CIC can be switched on/off via magnetic-field-induced spin canting and exhibit remarkable directional reversibility upon the reversal of the spin-canting direction. Furthermore, they demonstrate significant sensitivity to both the spin-canting angle and spin-orbit coupling (SOC). Such features clearly distinguish them from counterparts stemming from noncentrosymmetric crystal structures or collinear magnetic orderings. Moreover, we find that both the CSC and CIC in the finite-energy regime slightly above the band edge are governed by an unusual optical transition channel involving deeper valence bands, distinct from other optical responses in CrSBr \cite{SA,zhu_2024_CrSBr_nanoletter_review}.

\section*{Results}\label{Results}
\subsection*{Classification of BPVE}
As the BPVE encompasses various second-order photocurrents with distinct mechanisms, symmetry requirements and physical origins, a comprehensive classification is crucial. The total rectification current generated by the BPVE can be expressed as \cite{Qian-2020-npjCM-MIC&NSC_in_MnBi2Te4,Chen_2022}
\begin{equation}
\begin{split}
    J^a = & \left[ \sigma_{(i)}^{abc} + \tau\eta_{(c)}^{abc} \right] \text{Re} \left\{ E^b(\omega)E^c(-\omega) \right\} \\
          & + \left[ \sigma_{(c)}^{abc} + \tau\eta_{(i)}^{abc} \right] \text{Im} \left\{ E^b(\omega)E^c(-\omega) \right\},
\end{split}
\label{rectification}
\end{equation}
where $\sigma$ and $\eta$ represent the photoconductivities of the shift current and injection current, respectively (the microscopic formulas are detailed in Section IB of Supporting Information \cite{SI}). $E(\omega)$ is the electric field of the incident light, with $a$, $b$, and $c$ representing the Cartesian coordinates. The Einstein summation convention is adopted in all formulas for repeated indices. $\tau$ denotes the current-relaxation time \cite{Yan-2020-prr-carrier_lifetime}.
Equation~\ref{rectification} encapsulates a classification of the BPVE based on three aspects. First, the dependence on $\tau$ distinguishes the injection current mechanism from the shift current. Unlike the instantaneous shift current, the injection current grows linearly with time under illumination before reaching a steady state determined by $\tau$. Second, the terms associated with the real part of the electric field product can be excited by either linearly polarized light (LPL) or circularly polarized light (CPL) and are categorized as the linear photogalvanic effect (LPGE). Conversely, the terms associated with the imaginary part of the electric field can only be excited by CPL and are referred to as the circular photogalvanic effect (CPGE) \cite{Weng-2023-Quantum-Frontiers-MIC&MSC_in_EuSn2As2}. Here, we use a convention in which the imaginary prefactor commonly appearing in the CPGE term \cite{Chen_2022} is absorbed into $\sigma_{(c)}^{abc}$ and $\eta_{(i)}^{abc}$ in Eq. \ref{rectification}, so that all photoconductivities below are real coefficients that directly determine the photocurrent.
Finally, the subscripts $(i)$ and $(c)$ denote their $i$-type and $c$-type nature. Intuitively, the $i$-type BPVE components have long been thought to originate from noncentrosymmetric crystal structures and are referred to as the normal shift current (NSC) and normal injection current (NIC) due to their widespread occurrence \cite{Qian-2020-npjCM-MIC&NSC_in_MnBi2Te4,Chen_2022,Zhou-2022-nl-MIC&NSC_in_VSe2}. In contrast, the $c$-type BPVE components are referred to as the magnetic shift current (MSC) and magnetic injection current (MIC) because of their magnetic origins \cite{Qian-2020-npjCM-MIC&NSC_in_MnBi2Te4}.

\begin{table}
    \centering
    \small 
    \renewcommand\theadfont{\small} 
    \begin{tabular}{ccc}
        \hline \hline
         \thead{LPGE\\($\mathcal{P}$ broken)}& \thead{Crystal structure\\origin} & \thead{Magnetic structure\\origin}\\ \hline
         \thead{$i$-type\\($\mathcal{PT}$ broken)} & NSC & CSC \\ \hline
         \thead{$c$-type\\($\mathcal{T}$ broken)}  &  N.A. & MIC \\ \hline \hline
    \end{tabular}
    \caption{Classification of the linear photogalvanic effect (LPGE). NSC, CSC, and MIC denote the normal shift current, chiral shift current, and magnetic injection current, respectively. N.A. stands for not applicable. $\mathcal{P}$, $\mathcal{T}$, and $\mathcal{PT}$ represent inversion, time-reversal, and space-time-reversal symmetries, respectively.}
\label{tab:LPGE}
\end{table}
\begin{table}
    \centering
    \small 
    \renewcommand\theadfont{\small} 
    \begin{tabular}{ccc}
        \hline \hline
         \thead{CPGE\\($\mathcal{P}$ broken)}& \thead{Crystal structure\\origin} & \thead{Magnetic structure\\origin}\\ \hline
         \thead{$i$-type\\($\mathcal{PT}$ broken)} & NIC & CIC \\ \hline
         \thead{$c$-type\\($\mathcal{T}$ broken)}  &  N.A. & MSC \\ \hline \hline
    \end{tabular}
    \caption{Classification of the circular photogalvanic effect (CPGE). NIC, CIC, and MSC denote the normal injection current, chiral injection current, and magnetic shift current, respectively. N.A. stands for not applicable. $\mathcal{P}$, $\mathcal{T}$, and $\mathcal{PT}$ represent inversion, time-reversal, and space-time-reversal symmetries, respectively.}
    \label{tab:CPGE}
\end{table}
\begin{figure*}
    \centering
    \includegraphics[width=0.9\linewidth]{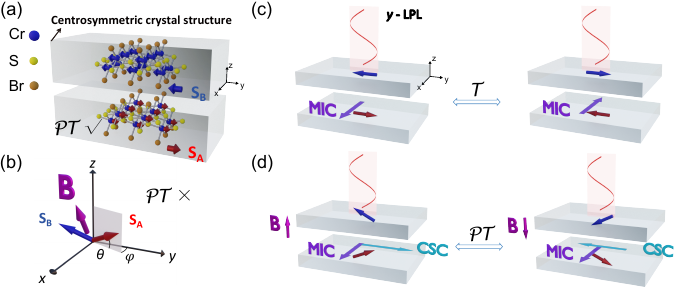}
    \caption{Concepts of LPGE in BL CrSBr. (a) Crystallographic and magnetic structures of $\mathcal{PT}$-symmetric BL AFM CrSBr. $\mathbf{S_A}$ and $\mathbf{S_B}$ represent the magnetic moments in the bottom and top layers respectively. (b) Definition of the spin-canting state and the generation of vector spin chirality in BL CrSBr with the magnetic field $\mathbf{B}$ in the $xz$ plane. This state breaks $\mathcal{PT}$ symmetry. (c) Schematic of the intrinsic MIC $\propto\eta_{\rm MIC}^{xyy}$ in BL AFM CrSBr under $y$-polarized light ($y$-LPL). The reversal of magnetic moments corresponds to a $\mathcal{T}$ operation and reverses the direction of the MIC. (d) Schematic of the emergence of the CSC $\propto\sigma_{\rm CSC}^{yyy}$ under $y$-LPL in BL cAFM CrSBr induced by a magnetic field $\mathbf{B}$ along the $z$ axis. Reversing $\mathbf{B}$ corresponds to a $\mathcal{PT}$ operation which switches the direction of the CSC while leaving the MIC unchanged.}
\label{Fig1}
\end{figure*}

\begin{table*}
    \centering
    \small 
    \begin{tabular}{ccccc}
        \hline \hline
         & (0$^\circ$,0$^\circ$) & ($\theta\neq 0^\circ$,0$^\circ$) & (0$^\circ$,$\varphi\neq 0^\circ$) & ($\theta\neq 0^\circ$,$\varphi\neq 0^\circ$) \\ \hline
         MPGs  & $m'mm$ & $m'2'm$ &   $22'2'$ & $2'$ \\ \hline
         $\sigma_{\rm CSC}^{abc}$  & N.A. & $yyy,yxx,xxy$ & N.A. & $yyy,yxx,xxy$ \\ \hline
         $\eta_{\rm MIC}^{abc}$ &  $xxx,xyy,yxy$ & $xxx,xyy,yxy$ & $xxx,xyy,yxy$ & $xxx,xyy,yxy$ \\
         \hline
         $\sigma_{\rm MSC}^{abc}$  & $yxy$ & $yxy$ & $yxy$ & $yxy$ \\
         \hline
         $\eta_{\rm CIC}^{abc}$  & N.A. & $xxy$ & N.A. & $xxy$ \\\hline \hline
    \end{tabular}
    \caption{Magnetic Point Groups (MPGs) of BL CrSBr with different spin-canting configurations and their corresponding nonzero BPVE photoconductivities. The notations of MPGs follow the direction sequence of $x$, $y$ and $z$. For LPGE the tensor components are symmetric with respect to the interchange of the last two indices satisfying $abc=acb$. In contrast, for CPGE the components are antisymmetric satisfying $abc=-acb$. Consequently redundant components are omitted. N.A. stands for not applicable.}
	\label{tab:sym}
\end{table*}
However, we find that the aforementioned classification is not exhaustive, and the $i$-type BPVE can be further classified. Fundamentally, the $i$-type nature of NSC and NIC originates from noncentrosymmetric crystal structures that simultaneously break $\mathcal{P}$ and $\mathcal{PT}$ symmetries \cite{sipe-2000-prb-over_band_gap_voltage,2018-Science-overcome_Shockley-Queisser_limit,2016-Shockley-Queisser-and-above-band-gap-photovoltage,Rappe-2012-prl-NSC_in_BaTiO3,Rappe-2012-prl-NSC_in_BaFeO3,Rappe-2016-prl-NSC_in_BiTeI&CsPbI3,Neaton-2017-prl-NSC_in_GeS,Moore-2017-NC-design_of_NSC,Qian-2019-SA-NSC&NIC_in_GeS,Huang-2023-prl-NSC_Heteronodal_Line_Systems,Kawasaki-2017-NC_NSC_experiment,Rappe-2017-npjCM_NSC_in_PbTiO3,Liu-2023-npjCM_2D_ferroelectrics,Yao-2024-NSC_in_PtBi2,Zhang-2024-nl-NSC_of_band_mixture,Zou-2024-NSC_in_GeTe}, whereas the $c$-type nature of MSC and MIC stems from magnetic structures that break $\mathcal{P}$ and $\mathcal{T}$ symmetries \cite{Yan-2019-NC-MIC_in_CrI3,Qian-2020-npjCM-MIC&NSC_in_MnBi2Te4,Li-2024-JPCL-MIC&NSC_Nb3I8,Li-2020-prb-MIC_in_MnBi2Te4,Zhou-2022-nl-MIC&NSC_in_VSe2}. However, a subtle case is often overlooked. In materials with centrosymmetric crystal structures, noncollinear spin structures can distinctively break $\mathcal{P}$ and $\mathcal{PT}$ symmetries akin to noncentrosymmetric crystal structures, effectively giving rise to $i$-type BPVE driven by noncollinear magnetism (details in Section IA of Supporting Information \cite{SI}). As these currents are closely tied to the vector spin chirality $\bm{\kappa}$ \cite{chiral_Hall_2020_prl,chiral_Hall_2021_communications_physics,chiral_photocurrents_2023_APL,SA}, we term them the CSC and CIC. We have summarized this detailed classification in Tables \ref{tab:LPGE} and \ref{tab:CPGE}. Despite having different physical origins, the CSC and CIC share identical symmetry properties and microscopic formulations with the NSC and NIC, and consequently both are described by $\sigma_{(i)}^{abc}$ and $\eta_{(i)}^{abc}$, whereas the MSC and MIC are described by $\sigma_{(c)}^{abc}$ and $\eta_{(c)}^{abc}$. It is worth noting that the CSC and CIC combine the $i$-type nature of the NSC and NIC with the magnetic origin of the MSC and MIC, suggesting unique properties that distinguish them from those traditional BPVE.

\begin{figure*}
    \centering
    \includegraphics[width=\linewidth]{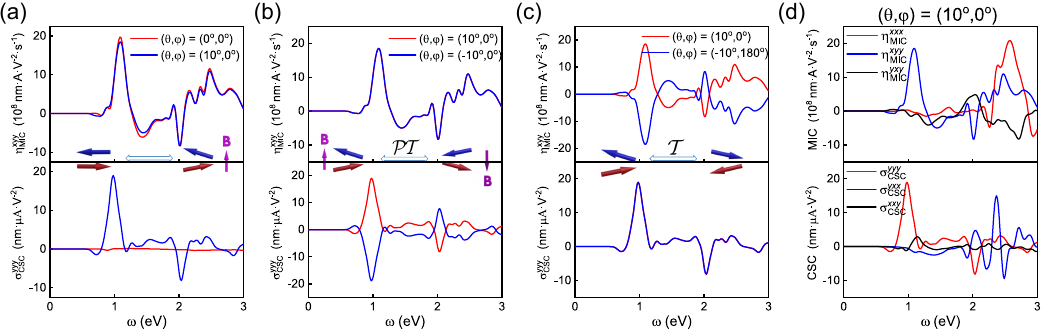}
    \caption{Generation and switching of LPGE in BL CrSBr. (a) Spin-canting-induced changes in $\eta_{\rm MIC}^{xyy}$ and $\sigma_{\rm CSC}^{yyy}$. The influence of reversing the spin-canting direction (b) and the influence of reversing the magnetic moments (c) on $\eta_{\rm MIC}^{xyy}$ and $\sigma_{\rm CSC}^{yyy}$. The insets are schematics of the corresponding processes. (d) Whole MIC and CSC photoconductivities at the (10$^\circ$, 0$^\circ$) canting state.
    } \label{Fig2}
\end{figure*}
\subsection*{CSC and CIC in BL CrSBr}
The newly discovered 2D van der Waals antiferromagnet CrSBr serves as an ideal platform to explore CSC and CIC. As shown in Fig. \ref{Fig1}(a), BL CrSBr has a centrosymmetric orthorhombic lattice with $D_{2h}$ crystal symmetry, where the magnetic moments in the bottom and top layers, $\mathbf{S_A}$ and $\mathbf{S_B}$, are aligned along the $y$-axis.
This A-type AFM ordering breaks both $\mathcal{P}$ and $\mathcal{T}$ symmetries but preserves the $\mathcal{PT}$ symmetry.
This feature results in the sole existence of $c$-type BPVE such as MSC and MIC and the vanishing of $i$-type BPVE such as NSC and NIC.
However, when BL CrSBr is driven into canted AFM (cAFM) states by an external magnetic field $\mathbf{B}$, the $\mathcal{PT}$ symmetry is broken due to the emergence of $\bm{\kappa} = \mathbf{S_A} \times \mathbf{S_B}$, activating nonzero CSC and CIC.
Figure \ref{Fig1}(b) shows a general case where $\mathbf{B}$ is perpendicular to the N\'eel vector (in the $xz$-plane). Under this condition, the corresponding canting states of BL CrSBr can be expressed in terms of the canting angles of $\mathbf{S_A}$, denoted as ($\theta, \varphi$). Here, $\theta \in [-90^\circ, 90^\circ]$ is the polar angle relative to the $xy$-plane, and $\varphi \in [0^\circ, 360^\circ)$ is the azimuthal angle with respect to the positive $y$-axis.
The corresponding magnetic point group and nonzero BPVE photoconductivity tensor components at different canting angles are summarized in Table~\ref{tab:sym}.
It is worth noting that $\theta \neq 0^\circ$ is a necessary condition for the existence of the CSC and CIC.
This is because the coexistence of the $y$-component of the N\'eel vector $L_y$ and the $z$-component of the net magnetization $M_z$ is essential for breaking the $\mathcal{PT}$, $\mathcal{C}_{2x}$, $\mathcal{M}_y$ and $\mathcal{C}_{2z}\mathcal{T}$ symmetries. This symmetry breaking in turn induces nonzero CSC components $\sigma_{\rm CSC}^{yyy}$, $\sigma_{\rm CSC}^{yxx}$, and $\sigma_{\rm CSC}^{xxy}$ as well as the CIC component $\eta_{\rm CIC}^{xxy}$.
Moreover, the $\mathcal{C}_{2y}\mathcal{T}$ symmetry present in all canting states listed in Table~\ref{tab:sym} consistently separates the $i$-type and $c$-type BPVE photoconductivity tensor components.
Based on these symmetry considerations, the total rectification current for BL cAFM CrSBr in Eq. \ref{rectification} can be written as
\begin{equation}
\begin{split}
    J^a = & \left[ \sigma_{\rm CSC}^{abc} + \tau\eta_{\rm MIC}^{abc} \right] \text{Re} \left\{ E^b(\omega)E^c(-\omega) \right\} \\
          & + \left[ \sigma_{\rm MSC}^{abc} + \tau\eta_{\rm CIC}^{abc} \right] \text{Im} \left\{ E^b(\omega)E^c(-\omega) \right\}.
\end{split}
\label{CrSBr_BPVE}
\end{equation}

Given that the LPGE can be generated by both LPL and CPL, rendering it more experimentally accessible, we focus our primary attention on the LPGE of BL CrSBr.
As depicted in Fig. \ref{Fig1}(c), in the pristine $\mathcal{PT}$-symmetric AFM phase under $y$-polarized LPL ($y$-LPL), only the MIC along the $x$-axis ($J_{\rm MIC}^x \propto \eta_{\rm MIC}^{xyy}$) is generated. Here, a global reversal of the magnetic moments constitutes a $\mathcal{T}$ operation which consequently reverses the sign of the MIC.
However, when the spins are canted by an external magnetic field $\mathbf{B}$ (e.g., along the $z$-axis), an additional CSC along the $y$-axis ($J_{\rm CSC}^y \propto \sigma_{\rm CSC}^{yyy}$) emerges as illustrated in Fig. \ref{Fig1}(d). In addition to the on/off switching capability, the CSC in BL CrSBr exhibits a unique directional reversibility driven by the spin-canting direction. This feature is distinct from both MIC and NSC. Notably, reversing the spin-canting direction is equivalent to a $\mathcal{PT}$ operation. This operation reverses the direction of the CSC while leaving the MIC unchanged, as illustrated in Fig. \ref{Fig1}(d).
This magnetic controllability distinguishes the CSC from the conventional NSC \cite{sipe-2000-prb-over_band_gap_voltage,2018-Science-overcome_Shockley-Queisser_limit,2016-Shockley-Queisser-and-above-band-gap-photovoltage,Rappe-2012-prl-NSC_in_BaTiO3,Rappe-2012-prl-NSC_in_BaFeO3,Rappe-2016-prl-NSC_in_BiTeI&CsPbI3,Neaton-2017-prl-NSC_in_GeS,Moore-2017-NC-design_of_NSC,Qian-2019-SA-NSC&NIC_in_GeS,Huang-2023-prl-NSC_Heteronodal_Line_Systems,Yao-2024-NSC_in_PtBi2,Zhang-2024-nl-NSC_of_band_mixture,Zou-2024-NSC_in_GeTe,Qian-2020-npjCM-MIC&NSC_in_MnBi2Te4,Zhou-2022-nl-MIC&NSC_in_VSe2,Li-2024-JPCL-MIC&NSC_Nb3I8}. Although the NSC can also be reversed by switching the ferroelectric polarization \cite{Yao-2024-NSC_in_PtBi2,Li-2024-JPCL-MIC&NSC_Nb3I8}, such reversal relies on a $\mathcal{P}$ operation rather than a $\mathcal{PT}$ operation and is practically more challenging due to the inherent rigidity of crystal structures.
Furthermore, the orthogonal spatial distribution of $J_{\rm MIC}^x$ and $J_{\rm CSC}^y$ under $y$-LPL offers excellent experimental feasibility. This configuration allows for the independent detection and control of MIC and CSC without ambiguity.

\begin{figure*}
    \centering
    \includegraphics[width=1\linewidth]{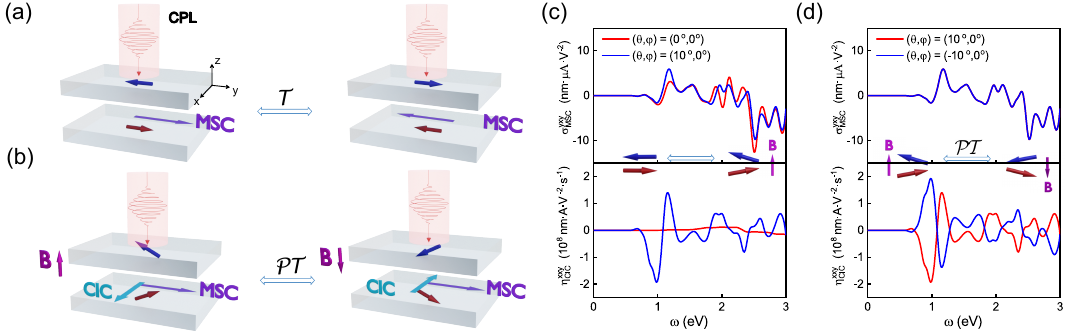}
    \caption{CPGE in BL CrSBr. (a) Schematic of the intrinsic MSC $\propto\sigma_{\rm MSC}^{yxy}$ in BL AFM CrSBr under circularly polarized light (CPL), and the influence of the $\mathcal{T}$ operation. The LPGE component excited by CPL is omitted in the schematic for simplicity. (b) Schematic of the emergence of the CIC $\propto\eta_{\rm CIC}^{xxy}$ under CPL in BL cAFM CrSBr induced by a magnetic field $\mathbf{B}$ along the $z$-axis, and the influence of the $\mathcal{PT}$ operation driven by the reversal of $\mathbf{B}$. (c) Spin-canting-induced changes in $\sigma_{\rm MSC}^{yxy}$ and $\eta_{\rm CIC}^{xxy}$. (d) The influence of reversing the spin-canting direction on $\sigma_{\rm MSC}^{yxy}$ and $\eta_{\rm CIC}^{xxy}$. The insets in (c) and (d) are schematics of the corresponding processes.}
\label{Fig3}
\end{figure*}

We performed first-principles calculations to verify the above symmetry analysis and quantify the CSC photoconductivities in BL CrSBr, taking MIC as a reference.
We took the out-of-plane canting state along $z$-axis with a small canting angle $\theta=10^\circ$ as an example since it can be easily achieved in experiments with a small $\mathbf{B}$ field of about 0.3\,T \cite{zhu_2021_nature-mater_CrSBr,zhu_2024_CrSBr_nanoletter_review}. In this configuration, the CSC components $\sigma_{\rm CSC}^{yyy}$, $\sigma_{\rm CSC}^{yxx}$, and $\sigma_{\rm CSC}^{xxy}$ emerge.
We examined the MIC and CSC under the illumination of $y$-LPL, i.e., MIC component $\eta_{\rm MIC}^{xyy}$ and CSC component $\sigma_{\rm CSC}^{yyy}$.
As shown in Fig. \ref{Fig2}(a), when the spin is canted toward the $z$-axis at $10^\circ$, the $\eta_{\rm MIC}^{xyy}$ component remains almost the same, while a significant $\sigma_{\rm CSC}^{yyy}$ component emerges, demonstrating the emergence of CSC induced by $\bm{\kappa}$.
It is worth noting that the magnitude of the CSC is comparable to that of the NSC in many other 2D magnets such as trilayer CrI$_3$ and monolayer H-VSe$_2$, which stems from noncentrosymmetric crystal structures, as shown in Fig. S2 \cite{SI}.
When the spin-canting direction is reversed, e.g., by reversing $\mathbf{B}$, $\eta_{\rm MIC}^{xyy}$ remains the same while $\sigma_{\rm CSC}^{yyy}$ reverses its sign, illustrating the switch of CSC direction, which is due to the $\mathcal{PT}$ operation, as shown in Fig. \ref{Fig2}(b).
In contrast, $\eta_{\rm MIC}^{xyy}$ reverses its sign while $\sigma_{\rm CSC}^{yyy}$ remains unchanged under $\mathcal{T}$-related canting states, i.e., ($10^\circ$, $0^\circ$) and ($-10^\circ$, $180^\circ$), demonstrating their $c$-type and $i$-type nature respectively, as shown in Fig. \ref{Fig2}(c).
These calculation results are in good agreement with our symmetry analysis above.
Furthermore, Fig. \ref{Fig2}(d) shows all independent in-plane tensor components of MIC and CSC photoconductivities.

Having established the properties of the LPGE in BL cAFM CrSBr, we now turn to the CPGE.
As illustrated in Fig. \ref{Fig3}(a), under CPL illumination, the CPGE response in the pristine $\mathcal{PT}$-symmetric AFM phase consists solely of the intrinsic MSC along the $y$-axis ($J_{\text{MSC}}^y \propto \sigma_{\text{MSC}}^{yxy}$).
It is worth noting that while the $c$-type response in the LPGE regime is an injection current (i.e., MIC), here in the CPGE regime, the $c$-type response manifests as a shift current (i.e., MSC).
Upon introducing spin canting via an external magnetic field $\mathbf{B}$, an additional CIC emerges along the $x$-axis ($J_{\text{CIC}}^x \propto \eta_{\text{CIC}}^{xxy}$) due to the breaking of $\mathcal{PT}$ symmetry by $\bm{\kappa}$, as shown in Fig. \ref{Fig3}(b).
This reveals an intriguing complementarity where the spin-chirality-driven BPVE manifests as a shift current under LPL (i.e., CSC) but switches to an injection current under CPL (i.e., CIC). Despite this distinct microscopic origin, the CIC shares the same $i$-type symmetry nature as the CSC.
Consequently, the CIC also exhibits the unique directional reversibility driven by the reversal of spin-canting direction, allowing it to be switched independently of the intrinsic MSC.
Note that the LPGE component excited by CPL is omitted in the schematic for simplicity.
We also performed first-principles calculations to confirm this symmetry analysis and quantify the CIC photoconductivities, taking the MSC as a reference. Figure \ref{Fig3}(c) demonstrates that while the intrinsic MSC component ($\sigma_{\text{MSC}}^{yxy}$) remains largely unchanged under a small spin-canting angle $(\theta,\varphi)=(10^{\circ},0^{\circ})$, a notable CIC component ($\eta_{\text{CIC}}^{xxy}$) emerges. Furthermore, as depicted in Fig. \ref{Fig3}(d), reversing the spin-canting direction reverses the sign of the CIC while leaving the MSC unchanged. This distinct switching behavior parallels that of the CSC and MIC, establishing a comprehensive framework for tunable spin-chirality-driven BPVE under both linear and circular optical excitations. The $i$-type and $c$-type nature of the CIC and MSC are also illustrated by their behavior under $\mathcal{T}$-related spin-canting states (see Fig. S3 \cite{SI}).

\begin{figure*}
    \centering
    \includegraphics[width=0.9\linewidth]{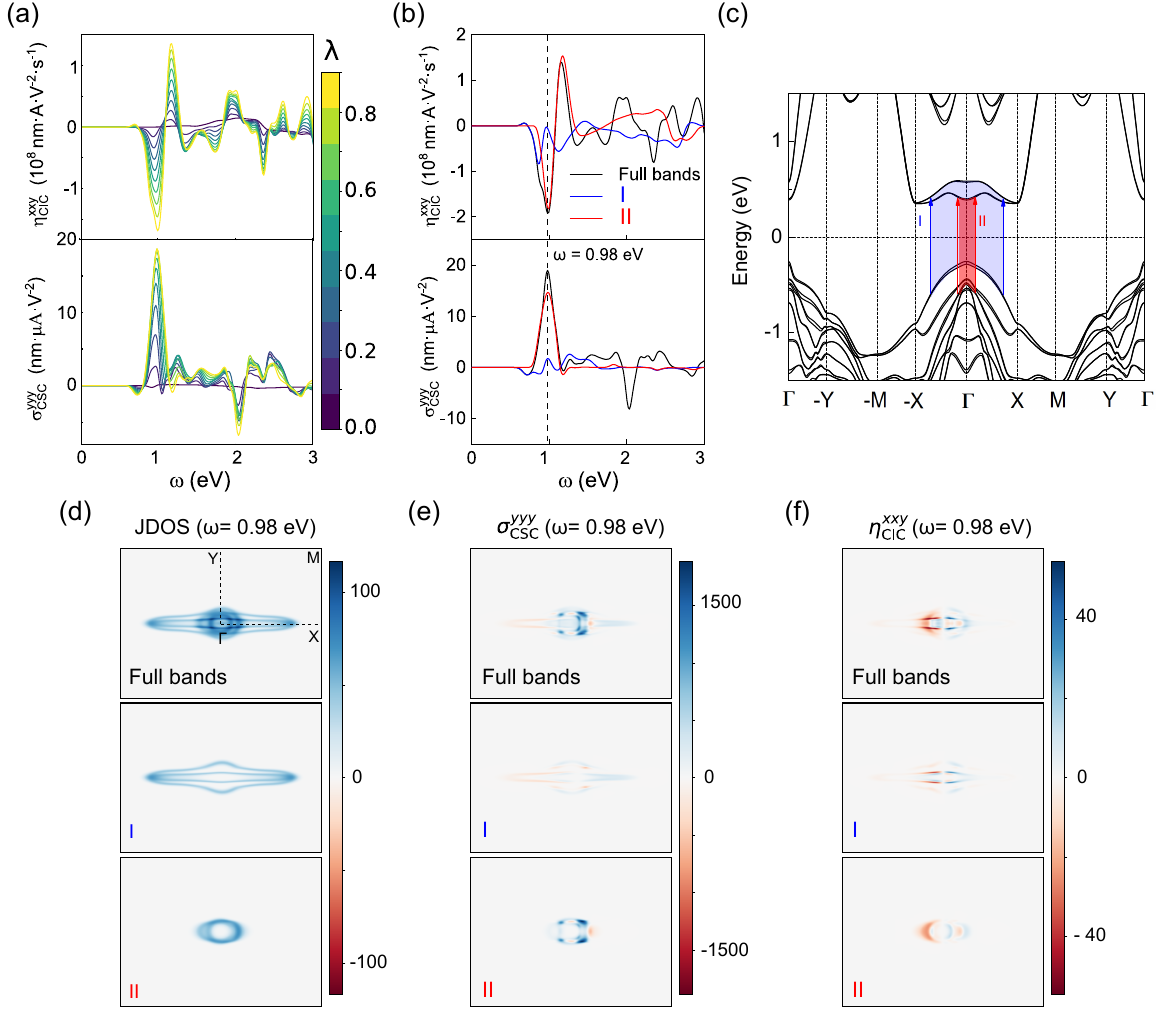}
    \caption{Mechanism of CSC and CIC in BL cAFM CrSBr exemplified by the $(\theta,\varphi)=(10^{\circ},0^{\circ})$ canting state. (a) The influence of SOC strength $\lambda$ on $\sigma_{\rm CSC}^{yyy}$ and $\eta_{\rm CIC}^{xxy}$. (b) Band-resolved analysis of the CSC photoconductivity $\sigma_{\rm CSC}^{yyy}$ and CIC photoconductivity $\eta_{\rm CIC}^{xxy}$. (c) Band structure of BL cAFM CrSBr. Channel I includes transitions from the highest two valence bands—the valence band maximum (VBM) and the band below it—to the lowest four conduction bands, which consist of the conduction band minimum (CBM) and the three bands above it. Channel II involves transitions from deeper valence bands, ranging from the second to the fifth band below the VBM.  (d)-(f) $\mathbf{k}$-resolved analysis corresponding to the peak at $\omega=0.98$\,eV showing the joint density of states (JDOS) in (d), CSC photoconductivity $\sigma_{\rm CSC}^{yyy}$ in (e), and CIC photoconductivity $\eta_{\rm CIC}^{xxy}$ in (f). The top, middle, and bottom panels represent the full band contribution, Channel I, and Channel II, respectively.}
\label{Fig4}
\end{figure*}

Having established the physical picture of both the LPGE and CPGE in BL cAFM CrSBr generated by LPL and CPL respectively, we now focus on the general driving mechanisms of the spin-chirality-driven components. As the interlayer cAFM ordering of BL CrSBr is essential for breaking $\mathcal{P}$ and $\mathcal{PT}$ symmetries to generate the CSC and CIC, we first explored the influence of SOC.
Taking the CSC component $\sigma_{\text{CSC}}^{yyy}$ and the CIC component $\eta_{\text{CIC}}^{xxy}$ as examples, Fig. \ref{Fig4}(a) illustrates the evolution of their spectra with varying SOC strength $\lambda$.
In the absence of SOC (i.e., $\lambda=0$), both $\sigma_{\text{CSC}}^{yyy}$ and $\eta_{\text{CIC}}^{xxy}$ vanish completely.
As $\lambda$ increases from 0 to 1, the intensities of both currents grow monotonically across the entire frequency spectrum. This indicates their magnetic origins and stands in stark contrast to the NSC and NIC. These conventional $i$-type BPVE are generally insensitive to magnetism \cite{Zhou-2022-nl-MIC&NSC_in_VSe2,Li-2024-JPCL-MIC&NSC_Nb3I8} and do not require SOC as a necessary condition \cite{Qian-2019-SA-NSC&NIC_in_GeS,Liu-2023-npjCM_2D_ferroelectrics}.
This SOC sensitivity also suggests that one can use heavier or lighter elements to explicitly tailor the CSC and CIC in AFM materials. The CSC and CIC of CrS$X$ ($X=$ Cl, Br, I) are shown in Fig. S4 \cite{SI}.
In addition to SOC, the interlayer coupling is another essential factor. The interlayer distance and layer thickness are crucial for the processes of symmetry breaking and the generation of CSC and CIC in BL CrSBr. We thus investigated their influence and found that they have a significant effect on both currents as shown in Figs. S5 and S6 \cite{SI}.

\begin{figure*}
    \centering
    \includegraphics[width=0.7\linewidth]{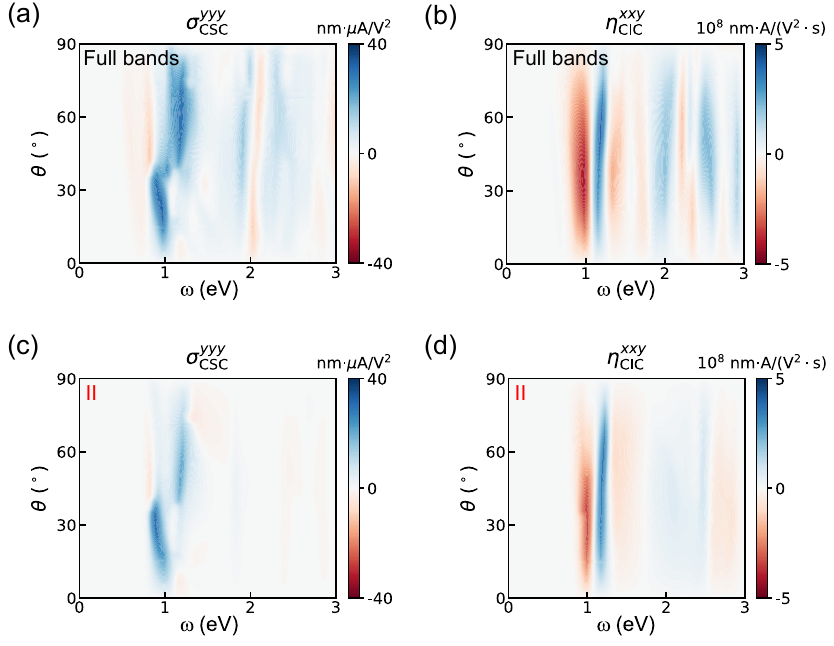}
    \caption{Continuous intensity modulation of CSC and CIC in BL cAFM CrSBr under different spin-canting angles. (a)-(b) Spectral evolution of the CSC photoconductivity $\sigma_{\rm CSC}^{yyy}$ and CIC photoconductivity $\eta_{\rm CIC}^{xxy}$ as a function of the photon energy $\omega$ and canting angle $\theta$ with $\varphi=0^{\circ}$. (c)-(d) The corresponding counterparts showing the isolated contributions from optical transition Channel II, as defined in Fig. \ref{Fig4}.}
\label{Fig5}
\end{figure*}
In addition to the above mechanisms, the first notable peaks of $\sigma_{\text{CSC}}^{yyy}$ and $\eta_{\text{CIC}}^{xxy}$ appear at the same energy of 0.98\,eV, suggesting a shared microscopic origin, as depicted in Fig. \ref{Fig4}(b). To elucidate this origin, we performed a band-resolved analysis focused on the prominent low-energy peak at $\omega = 0.98$\,eV.
As shown in Fig. \ref{Fig4}(c), the originally doubly degenerate bands of BL AFM CrSBr protected by $\mathcal{PT}$ symmetry are slightly split due to the small spin canting. The optical transitions around 0.98\,eV can be categorized into two distinct channels based on the participating bands. As indicated by the shaded regions in Fig. \ref{Fig4}(c), Channel I (blue shaded area) involves transitions near the band edges, while Channel II (red shaded area) involves transitions originating from deeper valence bands.
Channel I is composed of nearly parallel valence and conduction bands along the $\Gamma$-X path. This specific band geometry gives rise to the band nesting effect \cite{Huang_2023_NC_NbOCl,2020_nanolett_band-nesting,2013_prb_band-nesting,SA}. Consequently, this channel exhibits strong optical absorption and corresponds to the enhanced chiral SHG \cite{SA}.
The shared second-order nonlinear optical origin of BPVE and SHG hints that Channel I likely dominates the 0.98\,eV peak of the CSC and CIC.
However, our analysis reveals that the 0.98\,eV peak is dominated by Channel II instead of Channel I. As displayed in Fig. \ref{Fig4}(b), the calculated $\sigma_{\text{CSC}}^{yyy}$ and $\eta_{\text{CIC}}^{xxy}$ from Channel II almost perfectly reproduce the total CSC and CIC around the 0.98\,eV peak and the nearby spectral region. In contrast, the contribution from Channel I is negligible, which stands in sharp contrast to the dominant contribution from Channel I in chiral SHG in CrSBr \cite{SA}.
This unusual feature is further visualized by the $\mathbf{k}$-resolved quantities at $\omega=0.98$\,eV. As shown in Fig. \ref{Fig4}(d), the joint density of states (JDOS) shows contributions from both channels (definition of JDOS is detailed in Section IB of Supporting Information \cite{SI}). Channel I shows a broader area along $\Gamma$-X due to band nesting while Channel II shows a localized circle around the $\Gamma$ point. Consequently, the $\mathbf{k}$-resolved $\sigma_{\text{CSC}}^{yyy}$ and $\eta_{\text{CIC}}^{xxy}$ exhibit distributions similar to the JDOS as shown in Figs. \ref{Fig4}(e) and (f). However, for both CSC and CIC, the contributions of Channel I undergo strong sign cancellation in momentum space at $\pm\mathbf{k}$. Thus, they contribute little to the integrated spectrum. In contrast, Channel II contributes strong and non-canceled CSC and CIC in the $\mathbf{k}$-space. Therefore, it dominates the 0.98\,eV peak.
This finding indicates that the usually neglected optical transition channel involving deeper valence bands makes a significant contribution to these spin-chirality-driven BPVE. This distinguishes them from many other optical responses in CrSBr \cite{2020_AM_CrSBr,zhu_2021_nanoletters_CrSBr,zhu_2021_nature-mater_CrSBr,xu_2022_nature-nanotech_CrSBr,zhu_2022_nature_CrSBr,wang_2022_acsnano_CrSBr,xu_2023_nature-nanotech_CrSBr_tunable_exciton-magnon_coupling, xu_2025_nature-nanotech_CrSBr_tunable_exciton-magnon_coupling, 2023_Nature_CrSBr_tunable_MO,2024-NC-CrSBr-doping,zhu_2024_CrSBr_nanoletter_review,SA}.

Given the tunability of the magnitude of $\bm{\kappa}$, the associated CSC and CIC are expected to exhibit remarkable continuous modulation under different spin-canting angles. We thus varied the spin-canting angle $\theta$ to simulate the situation with an increasing magnetic field $\mathbf{B}$ applied along the $z$ axis in experiments, as depicted in Fig. S7 \cite{SI} (with a saturation field of $\sim 2$\,T) \cite{2020_AM_CrSBr,zhu_2021_nanoletters_CrSBr,zhu_2021_nature-mater_CrSBr,xu_2022_nature-nanotech_CrSBr,zhu_2022_nature_CrSBr,wang_2022_acsnano_CrSBr}.
Constrained by symmetry, the CSC and CIC are strictly forbidden in the pristine AFM state $(\theta, \varphi)=(0^\circ, 0^\circ)$ due to the presence of $\mathcal{PT}$ symmetry. Similarly, they vanish in the saturated ferromagnetic state $(\theta, \varphi)=(90^\circ, 0^\circ)$ owing to the restoration of $\mathcal{P}$ symmetry. Consequently, as $\theta$ increases from $0^\circ$ to $90^\circ$, the photoconductivities $\sigma_{\rm CSC}^{yyy}$ and $\eta_{\rm CIC}^{xxy}$ display a nonmonotonic evolution with a remarkably rich spectral landscape, as shown in Figs. \ref{Fig5}(a) and \ref{Fig5}(b). Specifically, their intensities initially grow, reach a maximum near $\theta = 45^\circ$, and subsequently decay to zero (other CSC tensor components are shown in Fig. S7 \cite{SI}). Notably, the prominent resonance peak of $\sigma_{\rm CSC}^{yyy}$ around $1.0$~eV undergoes a pronounced energy shift as the canting angle sweeps across $45^\circ$. Such strong sensitivity to the spin-canting angle allows the CSC and CIC intensities to be continuously and dynamically modulated by an external magnetic field and, conversely, enables them to serve as highly sensitive optical probes of the underlying magnetic ordering.
Crucially, as shown in Figs. \ref{Fig5}(c) and \ref{Fig5}(d), under varying spin-canting angles, the isolated contribution from Channel II consistently reproduces the main features of the total photoconductivity in the finite-energy regime around the 0.98\,eV peak. The individual spectra of $\sigma_{\rm CSC}^{yyy}$ and $\eta_{\rm CIC}^{xxy}$ presented in Fig. S8 \cite{SI} further illustrate this persistent behavior. Such dominance confirms that Channel II acts as a robust and dominant optical transition channel for generating $\sigma^{yyy}_{\mathrm{CSC}}$ and $\eta^{xxy}_{\mathrm{CIC}}$ in this finite-energy regime over the full range of spin-canting angles.
In contrast to the dramatic evolution of the photoconductivity, the JDOS remains insensitive to variations in $\theta$, as exemplified by the $\mathbf{k}$-resolved JDOS at $\omega=1.0$ eV in Fig. S9 \cite{SI}. This indicates that the giant modulation of the CSC and CIC is dominated by the sensitivity of microscopic quantities---specifically, the interband geometric quantity $I_{mn}^{abc}$ and the interband Berry curvature $\Omega_{mn}^{bc}$, as defined in Section IB of the Supporting Information \cite{SI}.

\section*{Discussion}\label{Discussion}
Our results suggest a direct experimental protocol for detecting spin-chirality-driven BPVE in BL CrSBr through photocurrent measurements below its N\'eel temperature $T_N\approx140$~K (with lower temperatures yielding more pronounced measured MIC in BL CrSBr \cite{Ideue-2026-Nat-Mater-MIC_in_CrSBr}), with an out-of-plane magnetic field $\mathbf{B}\parallel z$ maintained during illumination. For the representative $(\theta,\varphi)=(10^\circ,0^\circ)$ canting state considered here, a field of approximately 0.3~T is sufficient to induce the required spin canting \cite{zhu_2021_nature-mater_CrSBr}. Under $y$-LPL, the intrinsic MIC and the CSC flow along orthogonal in-plane directions, allowing $J^x_{\rm MIC}$ and $J^y_{\rm CSC}$ to be separately measured in a multiterminal device configuration. Reversing the magnetic field reverses the spin-canting direction and consequently the direction of $J^y_{\rm CSC}$, while leaving $J^x_{\rm MIC}$ unchanged. Under CPL, an analogous field-reversal protocol effectively separates the CIC from the intrinsic MSC. These symmetry-selective directional reversals provide a stringent fingerprint for the spin-chirality-driven BPVE, helping to unambiguously distinguish it from contact-induced, photothermal, or device-asymmetry-related photocurrents, which are not constrained by the same $\mathcal{PT}$-related switching rules. Furthermore, given that the intrinsic MIC in BL CrSBr has been recently observed in experiments \cite{Ideue-2026-Nat-Mater-MIC_in_CrSBr}, the experimental realization of the CSC and CIC in this material under an external magnetic field emerges as a feasible and compelling next step.
Combining the experimental illumination conditions reported for BL CrSBr \cite{Ideue-2026-Nat-Mater-MIC_in_CrSBr} with our calculated photoconductivities, we estimate maximum photocurrents of approximately $0.45$~nA for the CSC over the calculated photon-energy range of 0--3~eV. Taking the relaxation time of $\tau=0.2$~ps for the injection-type responses, guided by the recently measured MIC relaxation time  in 2D AFM MnPSe$_3$ \cite{Wu-2026-arxiv-MSC_exp}, yields a corresponding maximum of approximately $0.91$~nA for the CIC (see Section IIA of the Supporting Information for details~\cite{SI}). Our estimated CSC and CIC peak currents are compatible with the experimentally reported MIC of approximately $0.09$~nA in BL CrSBr \cite{Ideue-2026-Nat-Mater-MIC_in_CrSBr}, indicating the reasonableness of the predicted values and their experimental feasibility under suitable excitation conditions.
These detectability-focused estimates inherently depend on excitation energy, injection relaxation time, and local-field effects. Evaluating full device-level metrics, such as open-circuit voltage, output power, and power-conversion efficiency, requires further investigation.

In addition, quasiparticle and excitonic effects, which are particularly important in 2D semiconductors \cite{Louie-2021-pnas-exciton_shift_current_GeS,Louie-2024-nl-exciton_nlo_h-BN_MoS2,Eda-2026-acsnano-excitonic_shift_current_MoS2}, may quantitatively modify the predicted spin-chirality-driven BPVE. Quasiparticle self-energy corrections can increase the band gap and relevant interband transition energies, thereby blueshifting the corresponding spectral features \cite{Rappe-2020-prb-exciton_shift_current_BTO_SnSe}. Furthermore, electron--hole interactions can redistribute spectral weight \cite{Rappe-2020-prb-exciton_shift_current_BTO_SnSe,Nakamura-2024-NC-exciton_shift_current_CuI}, generate new excitonic resonances \cite{Louie-2021-pnas-exciton_shift_current_GeS,Louie-2024-nl-exciton_nlo_h-BN_MoS2,Nakamura-2024-NC-exciton_shift_current_CuI,Eda-2026-acsnano-excitonic_shift_current_MoS2}, including subgap photocurrent resonances associated with bound excitons \cite{Louie-2021-pnas-exciton_shift_current_GeS,Nakamura-2024-NC-exciton_shift_current_CuI}, and strongly modify---and in many cases enhance---the magnitudes of the photocurrent responses \cite{Louie-2021-pnas-exciton_shift_current_GeS,Louie-2024-nl-exciton_nlo_h-BN_MoS2,Nakamura-2024-NC-exciton_shift_current_CuI,Eda-2026-acsnano-excitonic_shift_current_MoS2}. At the level of symmetry, however, the major conclusions of our work remain robust as long as the equilibrium magnetic point group is preserved, including the symmetry-allowed CSC and CIC tensor components, their $i$-type character, and their reversal upon reversing the canting direction. A fully quantitative treatment of excitonic effects on the predicted CSC and CIC using a $GW$-BSE approach \cite{Louie-2021-pnas-exciton_shift_current_GeS,Louie-2024-nl-exciton_nlo_h-BN_MoS2,Eda-2026-acsnano-excitonic_shift_current_MoS2} is worth investigating in the future.

Moreover, the applicability of the CSC and CIC extends far beyond the prototype of BL cAFM CrSBr.
First, the observed proportionality of the CSC and CIC to SOC, along with their enhanced magnitudes in the CrS$X$ ($X=$ Cl, Br, I) series for heavier halogens, demonstrates that halogen substitution provides a general chemical route to tune SOC, thereby modulating the magnitude of the CSC and CIC across a broad range of halogen-containing materials. Additionally, the pronounced sensitivity of the CSC and CIC to interlayer coupling further identifies interlayer engineering as a complementary structural route for controlling their magnitudes, e.g., by experimentally exerting uniaxial pressure to reduce the van der Waals gap of layered magnets. Furthermore,
in contrast to the nonzero $\bm{\kappa}$ induced by an external magnetic field in BL cAFM CrSBr, many materials harboring native noncollinear spin architectures, including spin-spirals, frustrated noncollinear orders, and skyrmion lattices, naturally possess $\bm{\kappa}$, thereby providing a vast playground for exploring these chiral photocurrents. Across these diverse platforms, dynamically tuning $\bm{\kappa}$ via external stimuli like magnetic fields, epitaxial strain, or moir\'e interfacial engineering presents unprecedented opportunities for the active manipulation of the CSC and CIC.

\section*{Conclusion}\label{Conclusion}
In summary, we uncovered an unconventional type of spin-chirality-driven BPVE, which is $i$-type but originates from magnetism, including the CSC and CIC. Using BL AFM CrSBr as a prototype, we not only theoretically demonstrated the emergence of the CSC and CIC under magnetic-field-induced spin canting but also demonstrated their exceptional tunability. This includes spin-canting-induced on/off switching and directional reversibility governed by the reversal of the spin-canting direction, as well as remarkable intensity tunability under different spin-canting angles.
In general, we revealed that the driving mechanisms of these chiral photocurrents are SOC and interlayer coupling.
Specifically, we found that the CSC and CIC in the finite-energy regime slightly above the band edge of CrSBr are governed by an unusual optical transition channel involving deeper valence bands.
Beyond their high experimental accessibility, the underlying generality of our findings implies broad applicability to other 2D and 3D antiferromagnets with similar symmetries, as well as to more complex noncollinear magnetic architectures, opening vast avenues for future exploration.

\section*{Methods}\label{Methods}
 \subsection*{First-principles calculations}
First-principles calculations were carried out using the Vienna $Ab \ initio$ Simulation Package (VASP) \cite{VASP}, incorporating SOC effects. We employed the projector augmented-wave (PAW) method \cite{PAW} in conjunction with the Perdew-Burke-Ernzerhof (PBE) exchange-correlation functional \cite{PBE}. A Hubbard $U$ correction with $U_{\text{eff}}=3$\,eV was applied to the Cr $3d$ orbitals. The van der Waals interactions in the layered structures were treated using the DFT-D3 method with zero damping \cite{DFT-D3}. The kinetic energy cutoff for the plane-wave basis was set to 500\,eV. For structural relaxations, the convergence criteria for the total energy and ionic forces were set to $10^{-6}$\,eV and 1\,meV/\AA, respectively. The Brillouin zone integration was performed using an $11\times9\times1$ $k$-point grid, and a vacuum layer exceeding 15\,\AA\ was employed.
 \subsection*{BPVE calculations}
Upon obtaining the converged electronic ground state, we employed the Wannier90 code \cite{wannier90} to construct localized Wannier functions, which served as the basis for the tight-binding Hamiltonian and subsequent optical response calculations \cite{Wang_2017}. For each CrSBr layer, the projection involved 44 atomic-like orbitals (Cr-$d$ and S/Br-$p$). A broadening parameter of 0.05\,eV was applied to the Dirac delta function. We found that a $400\times 400\times1$ $\mathbf{k}$-mesh is sufficient for the converged photoconductivities. We used a $400\times 400\times1$ $\mathbf{k}$-mesh for the calculations of $\mathbf{k}$-resolved quantities. Since the thickness of a two-dimensional material is not well defined, sheet photoconductivities are used throughout this work: $\sigma=\sigma_{\rm bulk}L_z$ and $\eta=\eta_{\rm bulk}L_z$, where $\sigma_{\rm bulk}$ and $\eta_{\rm bulk}$ are the photoconductivities calculated using the lattice constant $L_z$ along the out-of-plane direction.


\section*{Supporting Information}
Symmetry analysis and microscopic formulas for the bulk photovoltaic effect; the supporting calculations of BPVE in CrSBr.

\section*{Data Availability Statement}
All the data and simulation codes in this study are available from the corresponding authors upon request.

\section*{AUTHOR CONTRIBUTIONS}
D.W., M.Y. and Y.X. conceived the project. D.W. carried out the numerical calculations and analyzed the data. D.W. and M.Y. discussed the results and wrote the manuscript with input from all other authors. M.Y., Y.X. and W.D. supervised the project.

\section*{Notes}
The authors declare no competing financial interest.


\section*{Acknowledgements}
This work was supported by the National Natural Science Foundation of China (Grant No. 12361141826), the National Key Basic Research and Development Program of China (Grant No. 2024YFA1409100), the Fundamental and Interdisciplinary Disciplines Breakthrough Plan of the Ministry of Education of China (Grant No. JYB2025XDXM408), the Basic Science Center Project of NSFC (Grant No. 52388201), the National Natural Science Foundation of China (Grant No. 12334003), the National Natural Science Foundation of China (Grant No. 12421004), the National Key Basic Research and Development Program of China (Grant No. 2023YFA1406400), the Innovation Program for Quantum Science and Technology (Grant No. 2023ZD0300500) and Beijing Key Laboratory of Quantum AI. The calculations were performed at National Supercomputer Center in Tianjin using the Tianhe new generation supercomputer.

\renewcommand{\bibfont}{\fontsize{9pt}{10.15pt}\selectfont}
\setlength{\bibsep}{-0.6pt}
\bibliography{ref}

\end{document}